%% file: ms.tex
\documentclass[iop]{emulateapj}  
 \usepackage[bookmarks=false,breaklinks,colorlinks,citecolor=blue,linkcolor=red]{hyperref} 

 \usepackage{etoolbox}

\makeatletter

\patchcmd{\NAT@citex}
  {\@citea\NAT@hyper@{%
     \NAT@nmfmt{\NAT@nm}%
     \hyper@natlinkbreak{\NAT@aysep\NAT@spacechar}{\@citeb\@extra@b@citeb}%
     \NAT@date}}
  {\@citea\NAT@nmfmt{\NAT@nm}%
   \NAT@aysep\NAT@spacechar\NAT@hyper@{\NAT@date}}{}{}

\patchcmd{\NAT@citex}
  {\@citea\NAT@hyper@{%
     \NAT@nmfmt{\NAT@nm}%
     \hyper@natlinkbreak{\NAT@spacechar\NAT@@open\if*#1*\else#1\NAT@spacechar\fi}%
       {\@citeb\@extra@b@citeb}%
     \NAT@date}}
  {\@citea\NAT@nmfmt{\NAT@nm}%
   \NAT@spacechar\NAT@@open\if*#1*\else#1\NAT@spacechar\fi\NAT@hyper@{\NAT@date}}
  {}{}

\makeatother
\usepackage{booktabs}
\usepackage{xspace}
\usepackage{apjfonts}
\usepackage{amssymb, amsmath} 
\usepackage{natbibspacing, natbib}
\usepackage{aas_macros} 
\usepackage{graphics,graphicx}
\newcommand{\pmOne}{\mbox{$^{-1}$}\xspace}

\newcommand{\mstar}{\mbox{$M_*$}\xspace}
\newcommand{\Msun}{\mbox{$M_{\odot}$}\xspace}

\newcommand{\Lsun}{\mbox{$L_{\odot}$}\xspace}
\newcommand{\LIR}{\mbox{$L_{\rm IR}$}\xspace}
\newcommand{\LFIR}{\mbox{$L_{\rm FIR}$}\xspace}
\newcommand{\kms}{km\,s$^{-1}$\xspace}

\newcommand{\E}[1]{$\times10^{#1}$}
\newcommand{\petm}[2]{$^{+#1}_{-#2}$}
\newcommand{\eq}{\,=\,}
\newcommand{\ssim}{\,$\sim$\,}
\newcommand{\pmm}{\,$\pm$\,}
\newcommand{\Fig}[1]{Figure~\ref{fig:#1}}

\newcommand{\Tab}[1]{Table~\ref{tab:#1}}
\newcommand{\Sec}[1]{\S\ref{sec:#1}}
\newcommand\tna{\,\tablenotemark{a}}

\newcommand\tnb{\,\tablenotemark{b}}
\newcommand\tnc{\,\tablenotemark{c}}

\newcommand{\ncode}[1]{{\sc #1}}

\newcommand{\z}{$z$\xspace}
\newcommand{\obs}{observations\xspace}

\newcommand{\highz}{high-$z$\xspace}

\newcommand{\SF}{star formation\xspace}

\newcommand{\SB}{starburst\xspace}

\newcommand{\apm}{APM 08279$+$5255\xspace}
\slugcomment{Accepted to ApJ}

\makeatletter
\renewcommand\normalsize{\@setfontsize\normalsize\@xpt{12.5}}
\makeatother

\shorttitle{SED Properties of the $z$\ssim3.9 AGN-Starburst System \apm}
\shortauthors{Leung et al.}

\begin{document}
\title{
Constraining the Active Galactic Nucleus and Starburst Properties of the IR-luminous Quasar Host Galaxy \apm
at Redshift 4 with {\it SOFIA}}

\author{T. K. Daisy Leung\altaffilmark{1, 2}}
\author{Christopher C.\ Hayward\altaffilmark{2}}
\author{Caitlin M. Casey\altaffilmark{3}}
\author{Johannes Staguhn\altaffilmark{4, 5}}
\author{Attila Kovacs\altaffilmark{6}}
\author{C.\ Darren Dowell\altaffilmark{7}}

\affil{\textsuperscript{1} Department of Astronomy, Space Sciences Building, Cornell University, Ithaca, NY 14853, USA; }
\email{tleung@astro.cornell.edu}
\altaffiltext{2}{Center for Computational Astrophysics, Flatiron Institute, 162 Fifth Avenue, New York, NY 10010, USA}
\altaffiltext{3}{Department of Astronomy, The University of Texas at Austin, 2515 Speedway Blvd, Austin, TX 78712, USA}
\altaffiltext{4}{Department of Physics \& Astronomy, Johns Hopkins University, Baltimore, MD 21218, USA}
\altaffiltext{5}{NASA Goddard Space Flight Center, Greenbelt, MD 20771, USA}
\altaffiltext{6}{Harvard-Smithsonian Center for Astrophysics, MS-78, 60 Garden St, Cambridge, MA 02138, USA}
\altaffiltext{7}{NASA Jet Propulsion Laboratory, California Institute of Technology, 4800 Oak Grove Dr., Pasadena, CA 91109, USA}

\begin{abstract}
We present far-IR photometry and infrared spectrum of
the $z$\eq3.9114 quasar/starburst composite system
\apm obtained using the {\it Stratospheric Observatory for Infrared Astronomy}
(SOFIA)/HAWC$+$ and the {\it Spitzer Space Telescope} Infrared Spectrograph (IRS).
We decompose the IR-to-radio spectral energy distribution (SED), sampled in 51 bands, using
(i) a model comprised of two-temperature modified blackbodies (MBB) and radio power-laws and (ii)
a semi-analytic model, which also accounts for emission from a clumpy torus.
The latter is more realistic but requires a well-sampled SED, as possible here.
In the former model, we find temperatures of
$T_{d}^{\rm warm}$\eq296\petm{17}{15}\,K
and
$T_{d}^{\rm cold}$\eq110\petm{3}{3}\,K
for the warm and cold dust
components, respectively.
This model suggests that the cold dust component dominates the FIR energy budget (66\%) but contributes
only 17\% to the total IR luminosity.
Based on the torus models, we infer an inclination angle of $i$\eq15\petm{8}{8}$\degr$ and
the presence of silicate emission,
in accordance with the Type-1 active galactic nucleus nature of \apm.
Accounting for the torus' contribution to the FIR luminosity,
we find a lensing-corrected star formation rate of SFR\eq$3075$\,$\times$\,$(4/\mu_L)$\,\Msun\,yr\pmOne.
We find that
the central quasar contributes 30\% to the FIR luminosity but dominates the total IR luminosity (93\%).
The 30\% correction is in contrast to the 90\% reported in previous work.
In addition, the IR luminosity inferred from the torus model is a factor of two higher.
These differences highlight
the importance of adopting physically motivated models to properly account for
IR emission in \highz quasars, which is now possible with {\it SOFIA}/HAWC+.
\end{abstract}
\keywords{infrared: galaxies --
          galaxies: high-redshift --
          galaxies: ISM --
          galaxies: evolution --
          galaxies: quasars --
          galaxies: starburst}

\section{Introduction}   \label{sec:intro}
Active galactic nucleus (AGN)-starburst galaxies at high redshifts are powerful probes of the early evolution
of black holes (BH) and galaxy assembly, and they may be the progenitors of present-day massive spheroidal galaxies.
Circumstantial evidence such as
the tight BH mass-host properties relations observed
in the nearby universe (e.g., $M_{\rm BH} - \sigma_*$ and $M_{\rm BH} - M_{\rm bulge}$) and the similar trends found between the
cosmic star formation rate (SFR) and BH accretion rate histories
suggest that the growth of supermassive BHs (SMBHs) and their host galaxies are tightly linked (and possibly
causally linked; see reviews by \citealt[][]{Alexander12a, Heckman14a, Madau14a}, and, e.g., \citealt{Sparre15a}
for evidence from simulations).
The leading SMBH-host co-evolution picture suggests that galaxies evolve via
hierarchical galaxy mergers, in which massive galaxies evolve from a starburst-dominated phase to a quasar-dominated phase
and finally settle as `red and dead' spheroidal galaxies
\citep[e.g.,][]{Sanders88a, Dimatteo05a, Hopkins06a, Hopkins08a, Treister10a, Narayanan10b, Page12a}.
Recent studies report that many of the most luminous AGNs at high redshift appear to be offset from the local BH-host relations,
where \highz AGNs appear to have
more massive black holes than expected given their host galaxy properties
(e.g., \citealt{Walter04a, Riechers08a, Riechers08b, Merloni10a, Wang13a}).
This offset suggests that AGNs and their host galaxies may not grow
``in tandem'', especially
during their peak phases of activity (\z\eq1$-$4).
Currently, we still lack a clear picture relating the interplay between AGN activity and \SF in their host galaxies
at early cosmic epochs, which is essential for understanding their formation and evolution.
In theory, useful insights can be gained by determining how the far-infrared (FIR) luminosity
(which depends on the AGN accretion rate and SFR), the interstellar medium (ISM) conditions, and morphology of AGN host galaxies
differ between the quasar phase ($z$\ssim1\,$-$\,4) and the present-day
epoch and between the different phases of the starburst-quasar evolutionary sequence
in the proposed SMBH-host co-evolution picture.

Previous studies indicate that $\sim$30\% of \highz quasars exhibit strong emission at (sub-)millimeter
wavelengths \citep[e.g.,][]{Chapman05a, Alexander08a, Coppin08a},
indicating large quantities of dust contributing to
IR luminosities of $>10^{\rm 13}$\,\Lsun.
The large IR luminosities in these quasars are commonly attributed to cold dust heated by their \SF,
corresponding to SFRs of $\gtrsim$1000\,\Msun\,yr\pmOne
(e.g., \citealt{Wang08a}).
One of the key challenges in improving our understanding of the AGN-host co-evolution picture
stems from the limited number of
comprehensive studies that have measurements covering the rest-frame mid-IR parts of the
dust spectral energy distributions (SEDs) of \highz AGN-starburst systems, with only a few detections attained with
{\it The Infrared Astronomical Satellite (IRAS)},
the {\it Spitzer Space Telescope},
and the {\it Herschel Space Observatory} \citep[e.g.,][]{Rowan-Robinson91a, Rowan-Robinson93a, Barvainis95a, Polletta08b}.
Thus, the fraction of the cold dust heated by young stars (and thus the SFR) in \highz quasar hosts remains
uncertain \citep[e.g.,][]{Omont03a, Carilli05a, Netzer14a}.
By probing the hot and warm dust heated by the AGN in \highz quasars, one can better understand the
co-eval growth of SMBHs and their host galaxies in their early stages of evolution.
In addition, determining the geometry of the dusty torus together with the host properties, one can investigate
whether dust-obscured and unobscured AGNs in the early Universe represent quasars at different stages of evolution \citep[see the review by][]{Netzer15a} or
whether they are simply a manifestation of different viewing geometries
under the well-known AGN unification framework \citep[i.e., Type-1 versus Type-2 AGN; e.g.,][]{Antonucci93a, Urry95a}.

In this paper, we demonstrate the new capabilities of the
{\it Stratospheric Observatory for Infrared Astronomy (SOFIA)} enabled by the new sensitive FIR instrument,
the High-resolution Airborne Wideband Camera$+$ (HAWC$+$; \citealt{Dowell13a, Harper18a}), together with ancillary IR photometry, to constrain the AGN contribution to the IR luminosity of
\apm.
The target is among the most apparently IR-luminous, gas-rich, strongly lensed AGN-starburst systems known to date, at a redshift of $z$\ssim3.91
(RA, Dec J2000\eq08$^{\rm h}$31$^{\rm m}$41$\fs$7, $+$52$\degr$45$^\prime$17\farcs5; \citealt{Egami00a, Weiss07a}).
Existing studies suggest that \apm is a rare starbursting quasar, in which both the AGN and host galaxy are
undergoing extremely active phases, with an SFR of 5000\,$\times$\,$(4/\mu_L)$
and an Eddington ratio of $\lambda_{\rm edd}$\eq0.4 \citep{Saturni18a}.
Since \apm is magnified by $\mu_{\rm L}$\ssim4$-$100 \citep{Egami00a, Riechers09a},
it has an apparent bolometric luminosity of $L_{\rm bol}$\eq7\E{15}\,\Lsun \citep{Irwin98a}
and IR luminosity of \LIR$=$\,10$^{\rm 15}$\,$\mu_L$\pmOne\Lsun,
making it one of the most ideal \highz galaxies for a demonstration study for {\it SOFIA}/HAWC$+$.
Due to its extreme brightness, it is one of the best-studied \highz objects and thus
is currently one of the rare cases with a rich set of photometry.
This enables more realistic SED modeling for
this source for the first time, motivating our work presented here.
Applying complex and realistic models
to \z$>$\,0 AGNs has been done previously, but mainly out to \z$\lesssim$\,2 \citep{Williams18a, Zhuang18a}.
While \citet{Leipski14a} and \citet{Lyu16a} use similar approaches to decompose the SEDs of
$z$\ssim6 AGNs, these sources have less complete SED coverage than \apm.
In most cases, their IR SEDs are constrained by less than ten data points.
By combining new {\it SOFIA}/HAWC+ photometry and a {\it Spitzer}/Infrared Spectrograph (IRS) spectrum with ancillary data covering multiple IR bands (\Tab{photometry}), we improve constraints on the IR SED of \apm
relative to previous work and model its SED with different AGN torus models to infer its torus properties.

This paper is structured as follows.
In \Sec{obs}, we summarize the observations and procedures used to reduce the data.
We also report ancillary data used in our analysis.
In \Sec{results}, we present the results.
In \Sec{sed}, we detail the SED modeling analysis.
Finally, we discuss the results and implication of our findings in \Sec{diss}, and we
summarize the main results and present our conclusions in \Sec{sum}.
Throughout this paper, we use a concordance
$\Lambda$CDM cosmology with
parameters from the WMAP9 results:
$H_0$ = $69.32$ \kms Mpc\pmOne, $\Omega_{\rm M}$ = $0.29$, and
$\Omega_{\Lambda}$ = $0.71$ \citep{Hinshaw13a}.


\section{New Observations and Ancillary Data}   \label{sec:obs}

\subsection{New Data: {\it Spitzer/IRS}}
Observations of \apm were carried out with the Infrared Spectrograph (IRS) onboard the {\it Spitzer Space Telescope}
on 2008 December 4 for a total observing time of 4.5\,hr (Program ID: 50784; PI: Riechers).
The first order of the long wavelength, low resolution module (LL1: 19.5$-$38.0\,$\micron$) was used, covering the rest-frame 6.7\,$\micron$ polycyclic aromatic hydrocarbons (PAH) feature.
These observations were obtained in spectral mapping mode,
wherein the target was placed at six positions along the slit.
The slit was positioned relative to a reference star using ``high accuracy blue peak-up''\footnote{https://irsa.ipac.caltech.edu/data/SPITZER/docs/irs/irsinstrumenthandbook/85/}.

The Basic Calibrated Data (BCD) files were produced by the {\it Spitzer}
pipeline, which includes ramp fitting, dark sky subtraction, droop correction, linearity correction, flat-fielding, and wavelength calibration.
Before spectral extraction, the two-dimensional dispersed frames
were corrected for ``rogue'' (unstable) pixels using \ncode{irsclean}, latent charge removal and residual sky subtraction\footnote{https://irsa.ipac.caltech.edu/data/SPITZER/docs/dataanalysistools/cookbook/23/}.
Sky frames at each map position is created by combining the corrected frames separately.
Uncertainty frames associated with the combined frames were produced based on the variance of each pixel as a function of time in the sky frames. 
One-dimensional spectra were extracted at the position of the continuum emission using the \ncode{(spice)} software. A final spectrum was obtained by averaging over all map positions.
Based on the residual sky background, the median noise of the final spectrum is 5.0\,mJy per
$\Delta\lambda$\eq0.169\,$\micron$ wavelength bin (observed-frame).

\subsection{New Data: {\it SOFIA}/HAWC$+$}
Observations of \apm with {\it SOFIA}/HAWC$+$ were performed on 2017 October 18 at an altitude of $\sim$43,000\,ft during cycle 5
(Program ID: 05\_0165; PI: Riechers).
A single pointing was used to observe at 53 (Band A), 89 (Band C), and 154\,$\micron$ (Band D).
Total Intensity observing in On-the-fly Mapping mode was used, producing a continuous telescope motion with 1.5\,min Lissajous scans,
covering (regions slightly smaller than) the standard HAWC$+$ fields-of-view (FoVs) of
2\farcm7\,$\times$1\farcm7 (Band A),
4\farcm2\,$\times$2\farcm6 (Band C), and
7\farcm3\,$\times$4\farcm5 (Band D).
Five scans corresponding to a total integration time of 333\,s were obtained in Band A, 12 scans totaling 870\,s in Band C, and 12 scans totaling to 936\,s in Band D. The scans were centered in the middle of the R0/T0 subarrays. As such, the maps are made from R0/T0 only.

The observations were performed at elevation angles between 30$-$37$\degr$.
No direct measure of Precipitable Water Vapor (PWV) was used in the calibration. Instead, the HAWC$+$ pipeline
uses a standard atmospheric model (the ATRAN model; \citealt{Lord92a})
to compute the predicted water vapor overburden at the zenith. The assumed zenith opacities are
$\tau_z$\eq0.11, 0.21, 0.24 for Bands A, C, and D, respectively.

The data were reduced and processed using the \ncode{crush}\footnote{http://www.sigmyne.com/crush/} software, in which an atmospheric correction based on the ATRAN model was
applied. For flux calibration, we applied a fixed
instrumental conversion factor to convert readout counts to janskys per pixel.
The conversion factor was determined by comparing the readout counts to the true fluxes of the primary and secondary flux calibrators, Uranus and Neptune, using the \citet{Bendo13a} temperature model.
In the post-processed images, the PSF FWHMs are 7\farcs1, 11\farcs0, and 19\farcs5 at 53, 89, and 154\,$\micron$, respectively.
After reduction, the pixel scales
are 1\farcs00, 1\farcs55, and 2\farcs75 for Bands A, C, and D, respectively.

\subsection{Ancillary Data}

In the analysis, we include photometry published in the literature, as listed in \Tab{photometry}. These include photometry
covering observed-frame NIR-to-radio wavebands.

We also report unpublished {\it Spitzer Space Telescope}/IRAC and MIPS photometry based on the Spitzer Enhanced Imaging Products (SEIP),
which are available through the NASA/IPAC Infrared Science Archive (IRSA).
The photometry in all IRAC bands reported in \Tab{photometry}
was extracted using a 3\farcs8 diameter aperture, with aperture corrections applied.
The MIPS photometry at 24\,$\micron$ was extracted through PSF-fitting to the high-S/N detection. We retrieved the {\it Spitzer} images from the Spitzer Heritage Archive.

We also report {\it Herschel}/PACS photometry, which are
tabulated in the {\it Herschel}/PACS Point source catalog (HPPSC; \citealt{Marton17a}).
PACS photometric maps were generated using the \ncode{JScanam} task within the
Herschel Interactive Processing Environment (HIPE version 13.0.0; \citealt{Ott10a}).
Sources were identified using \ncode{sussextractor},
and the flux densities were obtained by performing aperture photometry.

Extraction of the {\it Herschel}/SPIRE photometry at 250, 350, and 500\,$\micron$ (\Tab{photometry})
was performed using \ncode{sussextractor} within
HIPE version 15.0.1 on Level 2 SPIRE maps obtained from the Herschel Science Archive.
These maps were reduced and processed using the SPIRE pipeline version SPGv14.1.0
within HIPE, with calibration tree SPIRE\_CAL\_14\_3.
The \ncode{sussextractor} task estimates the flux density from
an image convolved with a kernel derived from the SPIRE beam. The flux densities measured by \ncode{sussextractor} were confirmed using the Timeline Fitter,
which performs photometry by fitting a 2D elliptical Gaussian to the Level 1
 data at the source position given by the output of \ncode{sussextractor}.
 The fluxes obtained from both methods are consistent within the uncertainties.

\input{table_photo}


\section{Results}    \label{sec:results}
\begin{figure*}[tbph]
\centering
\includegraphics[width=\textwidth]{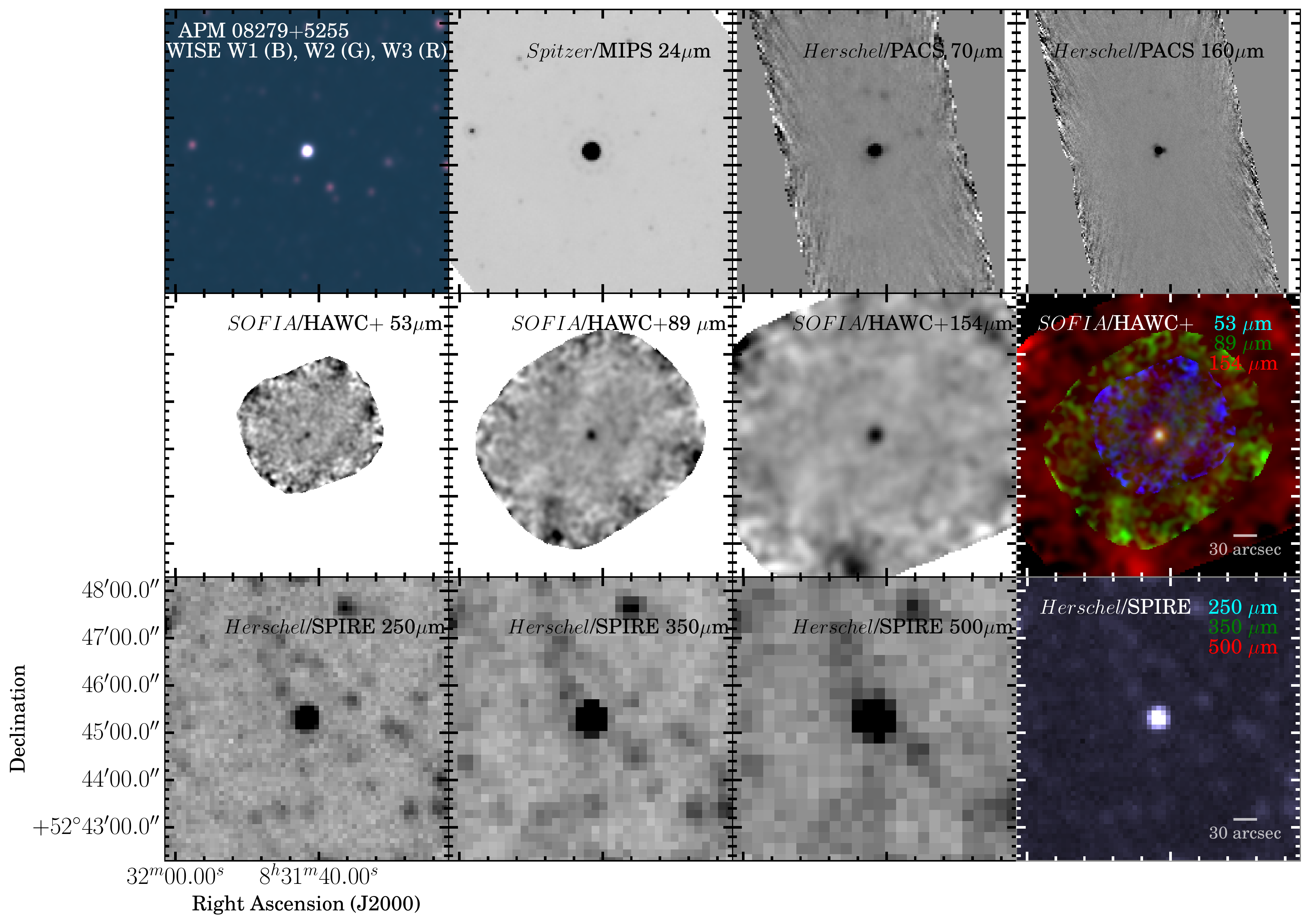}
\caption{Postage stamp images of \apm (from left to right).
Top row: {\it WISE} Bands W1, W2, and W3 RGB color composite;
{\it Spitzer}/MIPS 24\,$\micron$; and
{\it Herschel}/PACS 70 and 160\,$\micron$.
Middle row: {\it SOFIA}/HAWC+ 53, 89, and 154\,$\micron$ and color composite of the three
{\it SOFIA}/HAWC+ images.
Bottom row: {\it Herschel}/SPIRE 250, 350, and 500\,$\micron$ and color composite of the
three {\it Herschel}/SPIRE images.
\label{fig:cutout}}
\end{figure*}

We detect emission in all three bands covered by {\it SOFIA}/HAWC$+$ at high significance (\Fig{cutout}).
Point-source photometry were extracted using PSF-fitting and are listed in \Tab{photometry}.
Fitting a power-law to the IRS spectrum yields $S_\nu\propto\lambda^{0.01}$.

\begin{figure}[htbp]
\centering
\includegraphics[width=0.5\textwidth]{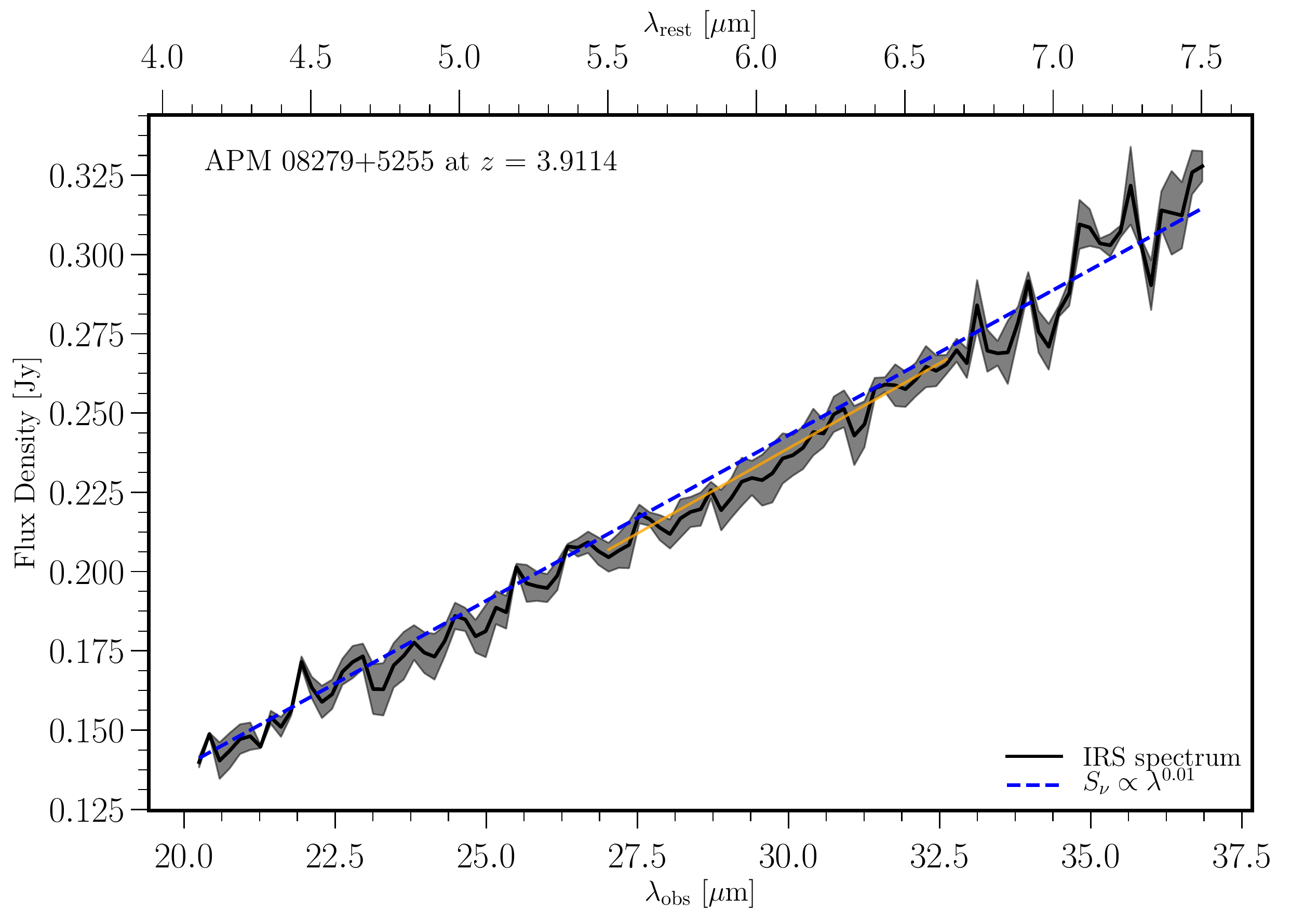}
\caption{
{\it Spitzer}/IRS spectrum of \apm. Shaded region shows the $\pm1\sigma$ noise level.
Orange line shows the power-law fit to the continuum covering rest-frame 5.5$-$6.7\,$\micron$.
Blue dotted line shows the best-fit power-law over the entire IRS spectrum.
\label{fig:irs}}
\end{figure}

The PAH feature at 6.2\,$\micron$ covered by our IRS \obs remains undetected down to a
$3\sigma$ limit of $<$3.4\E{-19}\,W\,cm$^{-2}$ (\Fig{irs}).
This limit was calculated by considering the residual flux density after removing the continuum
and assuming a linewidth covering rest-frame wavelengths from 5.9$-$6.5\,$\micron$. 
The continuum level was calculated by fitting a power-law over rest-frame wavelengths of 5.5$-$6.7\,$\micron$.

\section{Analysis: SED Modeling} \label{sec:sed}
\subsection{Analytic: Two-temperature MBB + Radio Power-laws}  \label{sec:mbb}
The IR-to-radio SED of \apm has been modeled previously by, e.g., \citet[][]{Rowan-Robinson00a},
\citet{Beelen06a}, \citet{Weiss07a}, and \citet{Riechers09a}; some of these models have their dust-emitting sizes ($r_0$) constrained through
large velocity gradient (LVG) modeling of the multi-$J$ CO and HCN line ratios
and spatially resolved CO imaging.
Existing models show that the observed dust SED can only be adequately described
by a model with {\em at least} two dust components --- a warmer and a cooler dust component.

\citet{Weiss07a} adopted a model assuming that the overall size of the dust-emitting region is similar to that of the CO line emission,
with relative area filling factors of $F_{\rm filling}$\eq0.75 and 0.25 for the warmer and cooler components, respectively,
and a dust absorption coefficient of
\begin{equation}
\kappa_\nu = \kappa_0 (\nu/\nu_0)^\beta,
\label{eqn:kappa}
\end{equation}
where $\kappa_0$\eq0.4\,cm$^2$/g, $\nu_0$\eq250\,GHz, and
the emissivity spectral index $\beta$ is fixed to $2$.
They find that the warmer component contributes $\sim$85\% to the FIR luminosity
(\LFIR\eq2\,$\times$\,10$^{14}$\,$\mu_L$\pmOne\,\Lsun) and suggest that this component is heated by the central AGN.
However, in these previous models,
the parameters describing the warm dust peak relied heavily on the constraints imposed by the {\it IRAS} measurement at
$\sim$100\,$\micron$, which has a $\sim$25\% uncertainty due to the limited S/N and the very large {\it IRAS} beam and pixel sizes.
Here, we update the SED by including our recently obtained {\it SOFIA}/HAWC$+$ data
and additional photometry (e.g., {\it Herschel}) obtained after the \citet{Riechers09a} study,
which extended the \citet{Weiss07a} study by also modeling the radio SED.
We note an outlier\footnote{This outlier is unlikely to result from contamination due to bright rest-frame optical lines, such as H$\alpha$, [N$\,${\sc ii}], [S$\,${\sc ii}], and [Ar$\,${\sc iii}], given that the broadband {\it WISE}/W1 flux would also be contaminated. }
at 3.5\,$\micron$ of $S_{\rm 3.5}$\eq27.3\pmm1.0\,mJy, which was obtained with the Palomar Telescope \citep{Egami00a}.

Following \citet{Riechers09a}, we first fit a simplified four-component model (two-temperature dust, free-free, and synchrotron)
to the rest-frame FIR-to-radio photometry, covering observed frame 53\,$\micron - 90$\,cm,
to illustrate how well the previously employed functional form fits all of the data obtained to date.
The functional form is expressed as follows:
\begin{equation}
S_\nu(\nu) = S_{\nu,\rm\,d}(\nu) + \sum_{i \rm\,=\,flat,\,steep} c_i \times \nu^{\alpha_i},
\label{eqn:dr09}
\end{equation}
where
$\nu$ is the frequency in GHz, $c_i$ is a coefficient, $\alpha_i$ is a spectral index, and $i$ represents
either the flat (free-free) or steep (synchrotron) power-law components.
The first term takes the functional form of a two-temperature MBB model, with each dust temperature component described by
\begin{equation}
S_\nu = [B_\nu (T_d) - B_\nu(T_{\rm CMB})] \frac{(1 - \exp^{-\tau_v})}{(1 + z)^3} \Omega_{\rm app},
\label{eqn:mbb}
\end{equation}
where $\Omega_{\rm app}$ is the apparent solid angle (the magnification factor is absorbed into this parameter).
The optical depth of each component is parameterized through the dust mass ($M_{\rm dust}$) and area filling factor ($F_{\rm filling}$):
\begin{equation}
\tau_\nu = \frac{\kappa_\nu~M_{\rm dust}}{(\pi~r_0^2)~F_{\rm filling}}.
\label{eqn:tau}
\end{equation}
The area filling factor clearly takes a value in the interval [0, 1]. We define $x\equiv M_{\rm dust}/r_0^2$ to reduce the extra dimension
and degeneracies between $M_{\rm dust}$ and $r_0$.
For the second term of Equation~\ref{eqn:dr09}, we adopt the best-fitting results from \citet{Riechers09a}
since no new constraints are added to the radio SED in this work.

We impose uniform priors on the model parameters as follows:
10\,K$<T_{d, \rm cold}< T_{d, \rm warm}$,
$T_{d, \rm warm}\leq500$\,K,
\mbox{F$_{\rm Filling}^{\rm cold}\in[0, 1]$}, \mbox{F$_{\rm Filling}^{\rm warm}\in[0, 1]$},
10$^{7}$\,\Msun\,pc$^{-2} < x_{\rm cold} < 10^{10}$\,\Msun\,pc$^{-2}$,
and 10$^4$\,\Msun\,pc$^{-2}< x_{\rm warm}<$10$^{10}$\,\Msun\,pc$^{-2}$.
We also impose the criterion that $x_{\rm warm} < x_{\rm cold}$.
These conditions are physically motivated based on previous results of \citet{Beelen06a} and \citet{Weiss07a}.
We find the best-fit model by sampling the posterior probability distributions (PDF) using a Markov Chain Monte Carlo (MCMC) approach.

\defcitealias{Riechers09a}{R09}
After discarding the first 2000 iterations in the burn-in
phase, we thin the chains via 40 iterations based on the auto-correlation time to obtain independent samples from the posterior PDFs,
yielding a best-fit model with $T_{d, \rm cold}$\eq110\petm{3}{3}\,K,
$T_{d, \rm warm}$\eq296\petm{17}{13}\,K,
\mbox{$F_{\rm Filling}^{\rm cold}$\eq0.9\petm{0.1}{0.1}},
$x_{\rm warm}$\eq2.5\petm{1.2}{1.0}\E{7}\,\Msun\,pc$^{-2}$, and
$x_{\rm cold}$\eq1.4\petm{0.1}{0.1}\E{9}\,\Msun\,pc$^{-2}$.
The best-fit values and their uncertainties are quoted based on the 16th, 50th, and 84th percentiles of the PDFs (See Appendix).
The best-fit SED is shown in \Fig{corner_t1_t2_ff1_ff2_md1_md2},
where we overplot the model presented by \citet{Riechers09a}\footnote{
The apparent difference in the \citetalias{Riechers09a} fit shown here compared to \citet{Riechers09a}
results from the fact that we adopt a source size of $r_0$\eq1150\,pc, following \citet[][cf. $r_0$\eq1300\,pc]{Weiss07a}.
The \citetalias{Riechers09a} fit here therefore
appears to undershoot the photometry at $\lambda_{\rm rest}$\ssim100\,$\micron$.
 \label{ftn:r09}
 }
 (hereafter, \citetalias{Riechers09a} fit) for comparison.

\begin{figure}[hptb]
\centering
\includegraphics[width=0.5\textwidth]{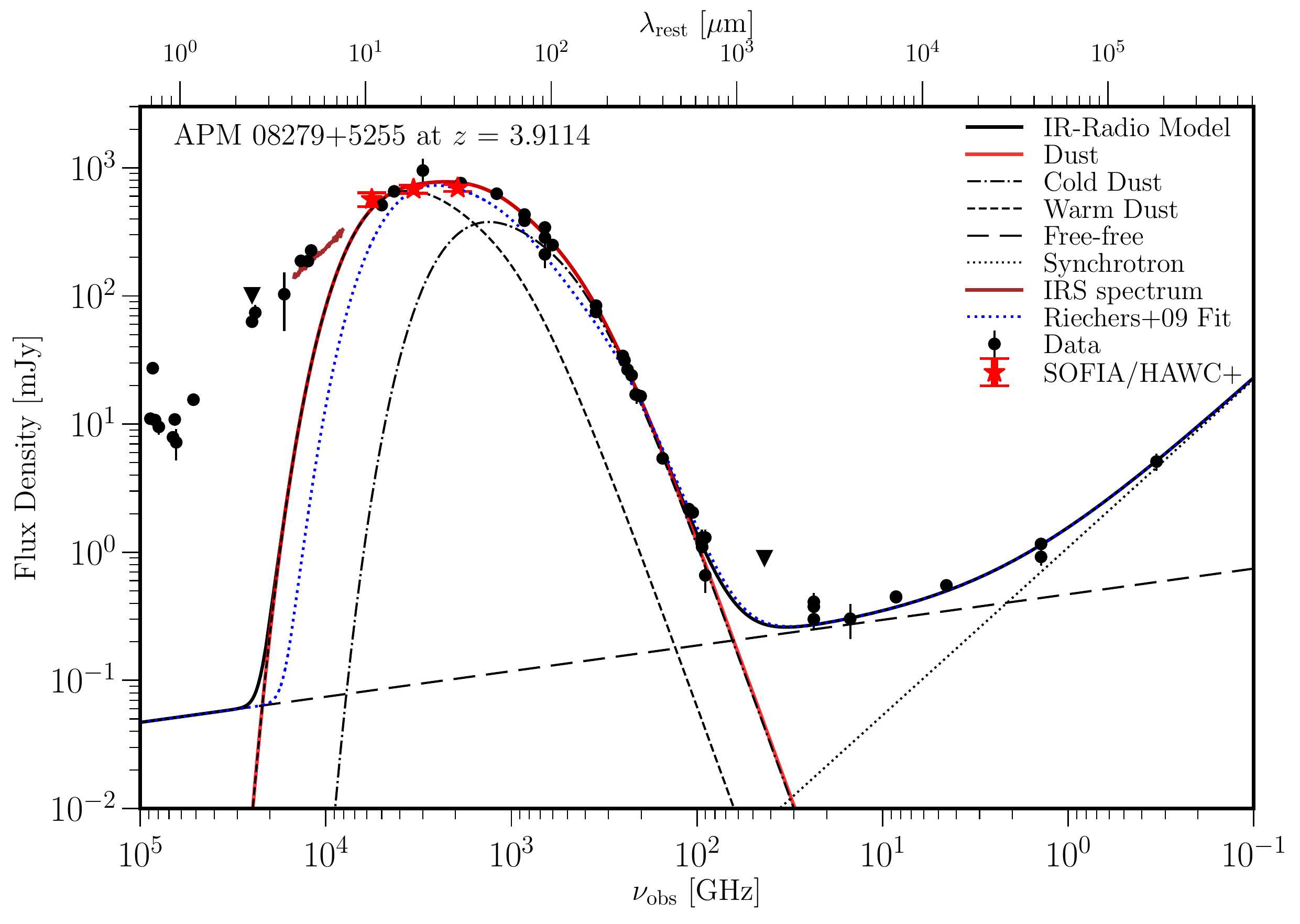}
\caption{
Ancillary photometry of \apm (black markers), our newly obtained {\it SOFIA}/HAWC$+$ data (red markers),
and the {\it Spitzer}/IRS spectrum (dark red line). The best-fit analytic model (solid black line) is overplotted.
Dashed-dotted and dashed lines show the SEDs of the warm and cold dust components, respectively.
Long-dashed and dotted lines show the radio free-free and synchrotron emission, respectively.
Previous model from \citetalias{Riechers09a} is overplotted (dotted blue line) for comparison; this
model underestimates the more comprehensive set of IR photometry presented in this work,
especially in the rest-frame wavelength range of $\lambda_{\rm rest}$\ssim1$-$100\,$\micron$
(see footnote~\ref{ftn:r09}).
\label{fig:corner_t1_t2_ff1_ff2_md1_md2}}
\end{figure}

\subsection{Semi-analytic: Including Clumpy Torus Models}     \label{sec:five}

Previous works suggest that \apm is a typical example of a co-existing starburst and quasar in a galaxy, where
its strong FIR luminosity due to cold dust emission has been attributed to the presence of a dust-obscured starburst, with a lensing-corrected
SFR of 5000\,$\times$\,$(4/\mu_L)$~\Msun\,yr\pmOne.    
In such AGN host galaxies, hot dust emission with a characteristic temperature of $\sim1500$\,K (e.g., \citealt{Barvainis87a, Deo11a})
due to reprocessed X-ray-to-optical AGN radiation is expected to peak in the NIR between 1 and 3\,$\micron$,
whereas warm dust emission from a dusty torus near the central AGN with characteristic temperature of $\sim400$\,K \citep{Schartmann05a, Burtscher15a}
should peak between 3 and 40\,$\micron$.
Based on the observed SED (\Fig{corner_t1_t2_ff1_ff2_md1_md2}), hot/warm dust emission due to a putative torus in \apm
is likely an important contributor to its IR emission (since its IR SED clearly peaks around rest-frame 20\,$\micron$).
We thus describe the SED of \apm (including the IRS spectrum)
using a physically motivated model, with
the UV-to-MIR part corresponding to emission from the accretion disk and and clumpy torus around it,
the FIR part described by a single-temperature MBB function,
and the radio part corresponding to free-free and synchrotron emission.
The cold dust and radio components are described using the same parameterization as Equations~\ref{eqn:dr09} and \ref{eqn:mbb}.
For the dust component, we assume an opacity defined using Equation~\ref{eqn:kappa}.
The spectral index $\beta$ is fixed to 2.0, following \citet{Priddey01a}, to minimize the number of free parameters.
For the AGN/torus component, we employ SED grids derived by performing
radiative transfer calculations of three-dimensional clumpy tori \citep{Honig10b}.
In this model, the optical depths and sizes of the dust clouds depend on their distance from the central source.
The clouds are randomly placed according to a spatial distribution function.
The torus SED is then calculated by summing the direct and indirect emission of the clumps.
More specifically, the model parameterizes the distribution of dust clouds in the tori using five parameters:
the radial dust-cloud distribution power-law index, $a$;
the half-opening angle, $\theta$ (describing the vertical distribution of clouds);
the average number of clouds along an equatorial line-of-sight, $N_0$;     
the total {\it V}-band optical depth, $\tau_V$; 
and the inclination angle, $i$.
The MIR SED is redder if more
dust clouds are distributed at larger distances.
Therefore, the power-law index $a$ is mostly constrained by the MIR photometry
(a more negative $a$ would imply a more compact dust distribution; see \citealt{Honig10a} for details).
The total number of clouds within the torus is thus related to $N_0$ through
\begin{equation}
N_{\rm tot} \propto \int N_0~\frac{\eta(r, z, \phi)}{R_{\rm cl}^2}~dV,
\end{equation}
where $\eta (r, z, \phi)$ describes the cloud distribution in the $r$-, $z$-, and $\phi$-space;
$R_{\rm cl}$ is the sublimation radius of a given cloud; and $V$ is the volume of the torus.
The torus models are scaled based on a luminosity factor ($L_{\rm scale}$), which is the luminosity of the accretion disk.
Physically, this determines the sublimation radius of the accretion disk/torus where dust clumps are distributed,
according to the following scaling properties:
\begin{equation}
\left(\frac{R_{\rm cl}}{0.5 \textrm{pc}}\right) \propto \left(\frac{L_{\rm scale}}{10^{46} \textrm{erg s}^{-1}}\right)^\frac{1}{2}\left(\frac{T_{\rm sub}}{1500~\textrm{K}}\right)^{-2.8},
\end{equation}
where $T_{\rm sub}$ is the sublimation temperature \citep[e.g.,][]{Barvainis87a}.
The numerical values in the denominators are the physical properties assumed in the model.
In the fitting process, we rescale the model flux to the distance of \apm.
We initialize $L_{\rm scale}$ based on the apparent X-ray luminosity (2$-$10\,keV; not lensing-corrected) of \apm,
which is $\log \left(L_X/L_\odot\right)$\eq49.4 \citep{Gofford15a}.

Since the \citet{Honig10a} templates are calculated based on discrete model parameters, we
interpolate the SED grids such that each parameter can be sampled as a continuous variable.
This is done in order to determine the PDFs of the parameters and thus obtain reliable uncertainties
on the model parameters. The parameters are interpolated in six dimensions (including the dimension of wavelength)
within the following boundaries:
$N_0\in$[0, 10], $a\in[-2.5,0.5]$, $\theta\in[0, 65]\degr$,
$\tau_V\in$[15, 90], and $i\in[0, 90]\degr$.

We place priors on the model parameters to ensure they are physically sensible.
We place flat priors of
$M_{d}^{\rm cold}\in[10^7, 10^{10}]$\,\Msun on the dust mass
and $T_{d}^{\rm cold}\in[30, 300]$\,K on the dust temperature.
The luminosity scaling factor is allowed to vary between $\log \left(L_{\rm scale}/L_\odot\right)\in[47, 49.4]$.
In addition to unobscured torus emission, stellar emission from the host galaxy may be significant
shortward of rest-frame $\sim 1 \, \micron$. Modeling the stellar emission would require introducing
multiple additional free parameters and properly accounting for potential differential magnification of
each of the components.
Thus, for simplicity, we fit the models to the photometry longward of 1\,$\mu$m only.

\begin{figure*}[!htbp]
\centering
\includegraphics[width=0.85\textwidth]{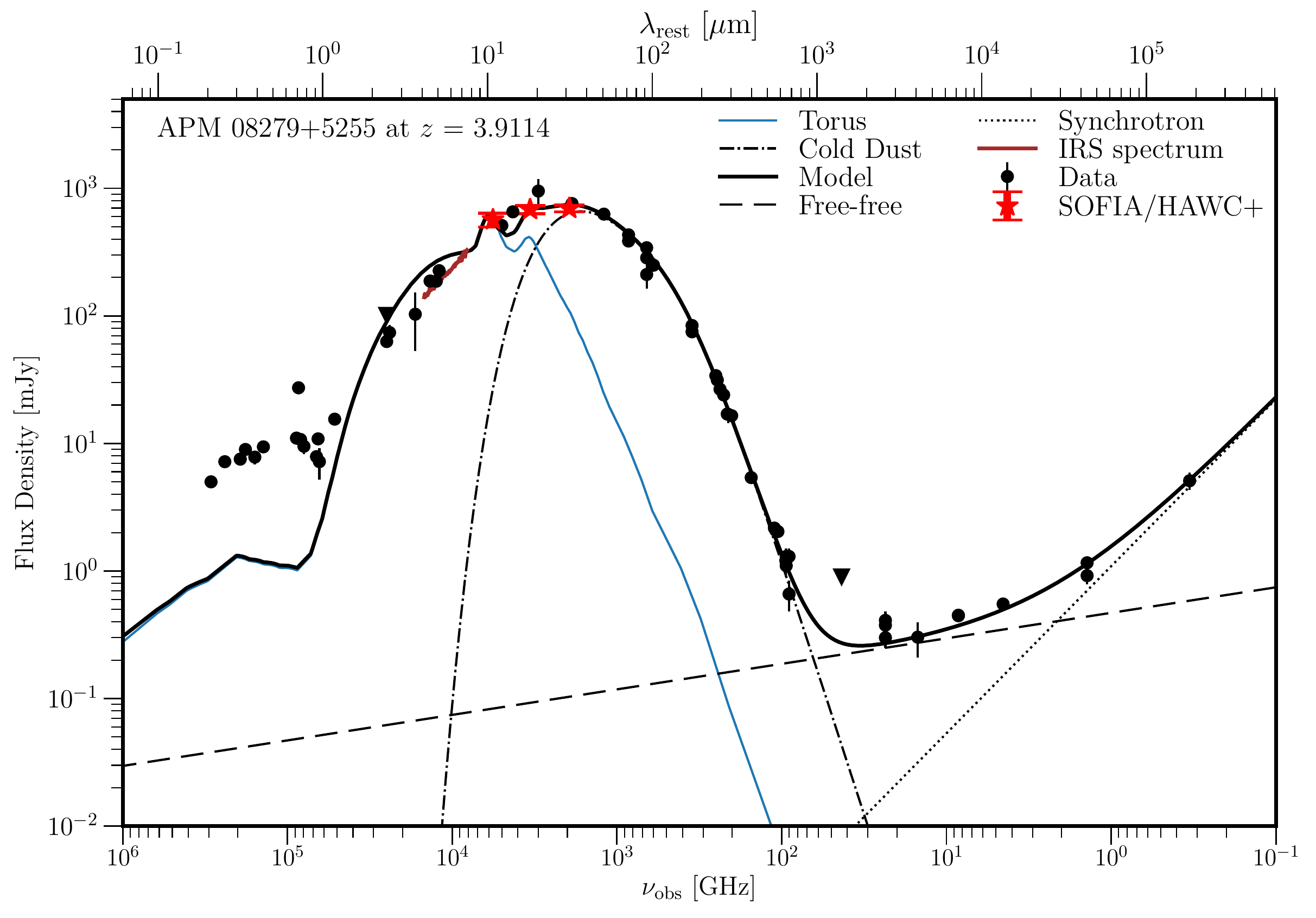}
\includegraphics[width=0.85\textwidth]{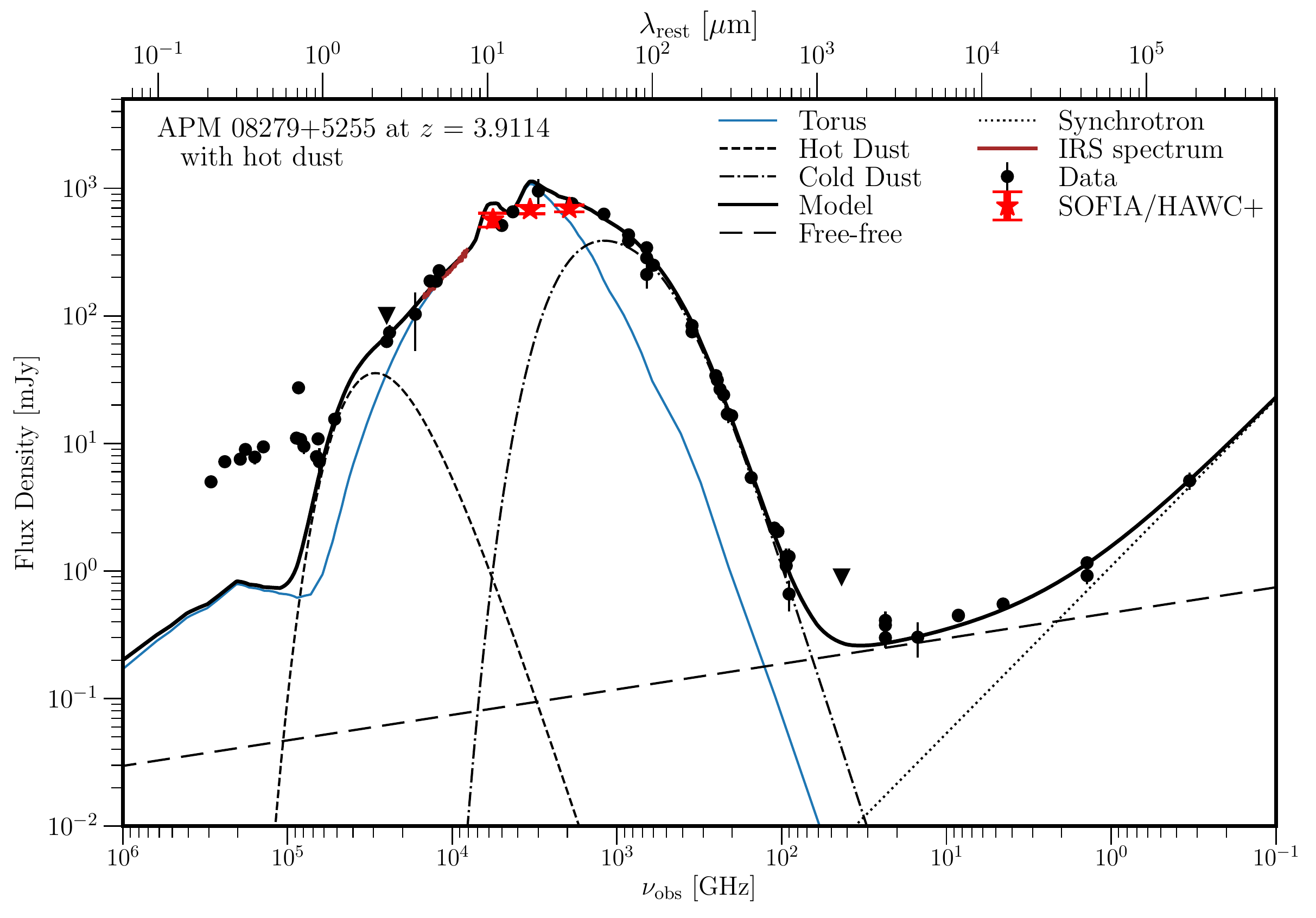}
\caption{
Solid black line shows the best-fit semi-analytic SED model fitted to the photometry at $\lambda>1$\,$\micron$ (black markers)
and the IRS spectrum (dark red line),
which covers rest-frame wavelengths $\lambda_{\rm rest}$\eq4.1\,$-$\,7.5\,$\micron$.
The new {\it SOFIA}/HAWC$+$ data are highlighted using red star symbols.
Top panel shows a model in which the UV-to-MIR part corresponds to emission emerging from an accretion disk
and a clumpy torus (solid blue line),
the FIR part is associated with cold dust emission from the host galaxy (dot-dashed line),
and the radio part is composed of free-free (dashed line)
and synchrotron emission (dotted line).
Bottom panel shows that a better fit is obtained by including an extra hot dust component (short dashed line).
The need for this extra hot dust component may suggest that \highz quasars have different dust compositions or/and geometries
than nearby systems. See \Sec{diss} for a more detailed discussion.
The two bumps near rest-frame 10$\,\micron$ correspond to the
silicate features of the model, which is constructed based on the nature of
nearby AGN systems (see \Sec{diss} for implication).
\label{fig:torus}}
\end{figure*}

The above model fails to yield a good fit to the NIR photometry ---
the torus component is skewed to the rest-frame NIR wavebands and over-predicts the fluxes observed in near- and mid-IR bands (top panel of \Fig{torus}).
Motivated by the work of \citet{Leipski14a},
we add an additional hot dust component to the overall model.
For this hot dust component, we place flat priors of
$r_{\rm hot}\in[10, 1300]$\,pc on its emitting radius,
$M_{d}^{\rm hot}\in[10, 10^5]$\,\Msun on its dust mass, and
$T_{d}^{\rm hot}\in[1000,1800]$\,K on its temperature.

After sampling the target posteriors using MCMC
with 100 walkers and 5\E{5} iterations, and after discarding the first 2\E{5} iterations
as the burn-in phase and thinning the chains by 50 iterations,
we identify the best-fit parameters together with the their uncertainties based on the
16th, 50th, and 84th percentiles of the PDFs.
The best-fit model is shown in the bottom panel of \Fig{torus}.
We summarize the best-fit parameters and the apparent luminosities integrated over the
total IR,
NIR (rest-frame 1$-$3\,$\micron$),
MIR (rest-frame 3$-$40\,$\micron$),
and FIR (rest-frame 42.5$-$122.5\,$\micron$) in the bottom section of \Tab{phy}.

We find an FIR luminosity of
$L_{\rm FIR}^{\rm Torus}$\eq4.7\petm{0.3}{0.5}\E{13}\,$\mu_L$\pmOne\,\Lsun for the torus component.
This corresponds to $\sim$30\% of the total FIR emission, which is in contrast to the 90\%
reported in previous work \citep{Weiss07a}.
Correcting for the torus contribution to the FIR emission,
we find an SFR of $3075$\,$\times$\,$(4/\mu_L)$\,\Msun\,yr\pmOne (assuming the \citealt{Chabrier03a} initial mass function).
This is consistent with that derived from the two-temperature MBB model, which is
SFR\eq2970\,$\times$\,$(4/\mu_L)$\,\Msun\,yr\pmOne.
We note that the IR luminosity derived from this model is
$\sim$2 times higher than that yielded by the fully analytic two-temperature dust model (\Sec{mbb}).
This discrepancy results mostly
from including a hot dust and a dusty torus component in the latter model, which is physically motivated
and better describes the IR SED of \apm (see Figure~\ref{fig:corner_t1_t2_ff1_ff2_md1_md2} and \ref{fig:torus}).
That is, the latter model describes the rest-frame near- and mid-IR photometry and the IRS spectrum,
whereas the former does not (by construction). This result highlights the importance of accounting for the NIR emission
using physically motivated models.

Including the near- and far-IR emission (i.e., covering rest-frame
1$-$122.5\,$\micron$), we find that dust emission due to the central AGN
accounts for $\sim$93\% of {\it the total IR luminosity}.
This is similar to the large fraction (83\%) inferred for the warm dust component in the fully analytic model.
Based on the latest lens model \citep{Riechers09a},
the magnification factor is found to vary by only a factor of $\mu_L$\ssim2$-$3 across a spatial extent of 20\,kpc.
Thus, differential magnification may cause the above percentage to be uncertain to a factor of 2$-$3.
However, if the cold dust is concentrated within the central 1\,kpc (see \citealt{Weiss07a} and \citealt{Riechers09a}),
the difference in the magnification factor between the AGN and more-extended cold dust would be $\lesssim$25\%.
With this uncertainty in mind, dust heating due to the central quasar appears to
be more significant than that from the \SF in the host galaxy of \apm,
and the fraction of the IR luminosity associated with the AGN is greater than those observed in
other \highz AGN/starburst composite systems studied to-date \citep[e.g.,][]{Pope08a, Coppin10a, Lyu16a}.

As noted above, the SED shortward of $\sim 1 \, \micron$ potentially has contributions from both the
AGN and stars in the host galaxy, and we have not attempted to fit the photometry shortward of 1\,$\micron$
for this reason.
If we assume that the residual flux density at 1\,$\micron$ unaccounted for by the torus model results from stellar emission in the
host galaxy,
we infer a stellar mass of $\mu_L M_*$\eq(3.6$-$5.3)\,$\times$\,10$^{13}$\,\Msun.
The range of stellar mass results from the
range of $H$-band mass-to-light ratio of $L_H/M_*$\eq$5.8-8.5$~(\Lsun$M_{\odot}^{-1}$) adopted \citep[e.g.,][]{Hainline11b},
which
corresponds to different population synthesis models (\citealt{BC03a} and \citealt{Maraston05a}) and
\SF histories (instantaneous \SF versus constant).
Taken at face value, the stellar mass of \apm is $\mu_L M_*\simeq$\,4\,$\times$\,10$^{13}$\Msun,
or $10^{13}$\,\Msun assuming a lensing magnification factor of $\mu_L$\eq4.
Such a stellar mass would place \apm among the most massive galaxies at \z$\sim$\,4 known.
Alternatively, the very large inferred stellar mass could suggest that the
lensing magnification for the stellar component is (at least) an order of magnitude higher than
the \citet{Riechers09a} value of $\mu_L \approx 4$ or that the residual flux
is not due entirely to stellar emission, perhaps because the torus models employed do not adequately
capture the geometry or/and dust properties of \apm.
Currently, the data at hand preclude us from favoring either scenario.

\input{table_out}

\section{Discussion}     \label{sec:diss}
In the case of the two-temperature dust SED model, we find that previous model consistently underestimates
the more comprehensive set of IR photometry included in this work, especially in the rest-frame
wavelength range covering $\lambda_{\rm rest}\in$[10, 100]\,$\mu$m
(\Fig{corner_t1_t2_ff1_ff2_md1_md2}),
with a maximum residual of 53\%.
The differences is mainly a direct consequence of the higher dust temperatures found here (110\,K vs. 65\,K and 296\,K vs. 220\,K).
In contrast to the results reported by \citet{Weiss07a},
we find that the cold dust component dominates the FIR luminosity of \apm ($\sim$\,66\%).
Regardless of these differences, we find comparable apparent total IR luminosities (rest-frame 1$-$1000\,$\micron$).
The (F)IR luminosity is summarized in the top section of \Tab{phy}.

When fitting the SED with clumpy AGN torus models in addition to the cold dust and radio
components, we find
a torus inclination angle consistent with the Type-1 nature of \apm (under the AGN unification model).
The best-fit model suggests the presence of silicate emission at rest-frame 10 and 18\,$\micron$\footnote{The 10\,$\micron$ feature is attributed to stretching of the Si-O bonds in silicates, whereas the 18\,$\micron$ feature is attributed to the O-Si-O bending mode.}, as commonly
observed in Type-1 AGNs with {\it Spitzer} \citep[e.g.,][]{Hao05a, Siebenmorgen05b, Sturm05a}.
If these features are real, their presence
is consistent with the AGN unification model, in which an AGN torus seen close to pole-on ($i$\eq0$\degr$) should exhibit
silicate emission features in the MIR (e.g., \citealt{Hao05a}; see also \citealt{Stalevski12a}),
as dust on the surface of the torus is heated to temperatures sufficiently hot to produce silicate emission features.
However, if these features are fictitious, this could be evidence that different torus models are needed to describe
AGN/\SB systems at high redshifts.

As mentioned in \Sec{five}, the best-fit semi-analytic model including the torus component without an extra
hot dust component overestimates the near-to-mid-IR fluxes. The MIR slope is also
inconsistent with the IRS spectrum.
We follow the work by \citet{Leipski14a} for \z$\sim$6 quasars and include an extra hot dust component in the model.
The extra hot dust component improves the fit and yields good agreement with the IR photometry.
However, the physical motivation for this extra hot dust component is unclear, as
the hottest dust emission should arise self-consistently in the clumpy AGN torus models
computed via radiative transfer, and there should be no need to add an additional ad hoc hot dust
component.
This need for an additional hot dust component may suggest that these powerful quasars in the early Universe
have very different dust compositions and distributions than the more-nearby systems
used to motivate AGN torus models. If this issue persists for larger samples of \highz AGN,
further improvements in AGN torus models would be warranted.

\section{Summary and Conclusions}     \label{sec:sum}
We observe the IR-luminous, strongly-lensed AGN-starburst system \apm at $z$\eq3.9114 using the
new sensitive FIR camera HAWC$+$ onboard {\it SOFIA} and the IRS onboard {\it Spitzer}.
While \apm is a Type-1 broad absorption line AGN, it clearly shows strong mid-IR emission
(even considering the viewing angle of \apm in the classical AGN unification model).
Hence, SED modeling tools such as \ncode{cigale} are inappropriate\footnote{
Modeling tools such as \ncode{cigale} expect Type-1 AGNs to emit predominately in UV and optical wavebands
and Type-2 AGNs to emit predominately in MIR wavelengths.
Since a large amount of dust is still present in \apm, the emission in the MIR waveband is
significantly underestimated in \ncode{cigale}
when one properly considers it as a Type-1 AGN.
} for \apm.
We combine the new data with ancillary data and
decompose the IR SED into hot dust, torus, and cold dust components
and the radio SED into thermal and non-thermal components (free-free and synchrotron).
To facilitate comparison with previous studies, we also fit simplified two-temperature modified blackbody models.

We constrain the structure of the putative dusty torus around the central AGN by fitting
3D radiative transfer models of clumpy tori.
Despite the uncertainties and degeneracies between the model parameters, some results are clear from the fit.
For instance, the best-fit inclination angle and the silicate features predicted by the best-fit model
are consistent with the Type-1 AGN nature of \apm,
which may be evidence supporting the unification model of AGN.

In the modified blackbody model presented in this work,
we find that the warm dust component contributes only $34\%$ of the far-IR luminosity.
Accounting for the torus' contribution to the FIR luminosity,
we find a lensing-corrected SFR of $3075$\,$\times$\,$(4/\mu_L)$\,\Msun\,yr\pmOne.
Differences compared to previous models are pointed out in \Sec{five}.

Accounting for emission due to the torus at near-to-far-IR wavelengths (rest-frame 1$-$122.5\,$\micron$),
we find that dust heating from the quasar contributes about 93\% to its {\em total} IR emission.
Therefore, the AGN plays an important role in driving the total dust luminosity of \apm, more so
than in other \highz AGN/starburst composite systems studied to date \citep[e.g.,][]{Pope08a, Coppin10a,
Lyu16a}.
Based on the residual flux density at 1\,$\micron$ unaccounted for by the torus model,
we find a stellar mass of \mstar$\simeq$\,$(4/\mu_L)$\E{13}\,\Msun.
At face value, assuming the \citet{Riechers09a} value of $\mu_L \approx 4$, the resulting stellar mass would place \apm among the most evolved and massive galaxies at \z$\sim$\,4
known, but it may also indicate that the true magnification is significantly greater than 4 and/or that the best-fit
AGN torus model is significantly underestimating the contribution of the AGN at $\sim$1\,$\micron$.

Our results demonstrate the use of the new and sensitive FIR camera onboard {\it SOFIA} to
constrain the physical properties of the dusty tori of quasars in the early Universe.
The differences between the simplified model and the more complex but physically motivated model
highlight the need for well-sampled IR SEDs of \highz quasars in order to disentangle
the effect of dust heating due to AGN versus \SF and thus to
more accurately determine the SFRs in AGN/starburst composite systems.
We note that the high IR luminosity of \apm may imply that our results are unlikely
to be representative of \highz AGNs.
The {\it Origins Space Telescope (OST)} will have the capabilities to characterize the torus and dust
components in \highz AGNs with a much broader range of properties, enabling
similar types of studies to be carried out for more typical and representative systems.

\acknowledgements

We thank the referee for the careful reading and for providing constructive comments on the manuscript.
The authors thank Dominik Riechers for his contributions to the {\it SOFIA}
and {\it Spitzer} proposals that resulted in observations used here, and for
access to data and data products used in this work.
The authors also
thank Lee Armus, Chris Carilli, Aaron Evans, Patrick Ogle, and Fabian
Walter for their role in this study, and Agatha Hodsman for providing
the reduction of the {\it Spitzer}/IRS data that is used in this work.
TKDL thanks Denis Burgarella for helpful discussion about the SED modeling code \ncode{cigale}.
TKDL acknowledges support from NASA through the New York Space Grant, award NAS2-97001 from the {\it SOFIA} MSO, and
support from the Simons Foundation.
TKDL acknowledges the hospitality of NASA's Goddard Space Flight Center and the Space Telescope Science Institute, where part of this manuscript was prepared.
The Flatiron Institute is funded by the Simons Foundation.
CMC thanks the University of Texas at Austin College of Natural Sciences, NSF grants AST-1714528 and AST-1814034, as well as support from a 2019 Cottrell Scholar Award from the Research Corporation for Science Advancement.
This work is based in part on observations made with the NASA/DLR {\it Stratospheric Observatory for Infrared Astronomy (SOFIA)}. {\it SOFIA} is jointly operated by the Universities Space Research Association, Inc. (USRA), under NASA contract NAS2-97001, and the Deutsches SOFIA Institut (DSI) under DLR contract 50 OK 0901 to the University of Stuttgart. Financial support for this work was provided by NASA through award \#SOF 05-0165 issued by USRA.
{\it Herschel} is an ESA space observatory with science instruments provided by European-led Principal Investigator consortia
and with important participation from NASA.
This work is based on observations made with the Spitzer Space Telescope, which is operated by the Jet Propulsion Laboratory, California Institute of Technology under a contract with NASA.
The IRS was a collaborative venture between Cornell University and Ball Aerospace Corporation funded by NASA through the Jet Propulsion Laboratory and Ames Research Center.
This research has made use of the NASA/ IPAC Infrared Science Archive, which is operated by the Jet Propulsion Laboratory, California Institute of Technology, under contract with the National Aeronautics and Space Administration.
Facilities: {\it SOFIA} (HAWC$+$), {\it Spitzer} (IRS, IRAC, MIPS), IRSA, {\it Herschel} (PACS, SPIRE)

\bibliographystyle{yahapj}
\bibliography{ref.bib}

\appendix
We show the 2D correlation plots and the 1D marginalized PDFs for the SED models presented in \Sec{mbb} and \Sec{five}.
The best-fit values and their uncertainties reported in \Tab{phy} are quoted based on the 16th, 50th, and 84th percentiles of the PDFs.

\begin{figure}[tbph]
\centering
\includegraphics[width=0.75\textwidth]{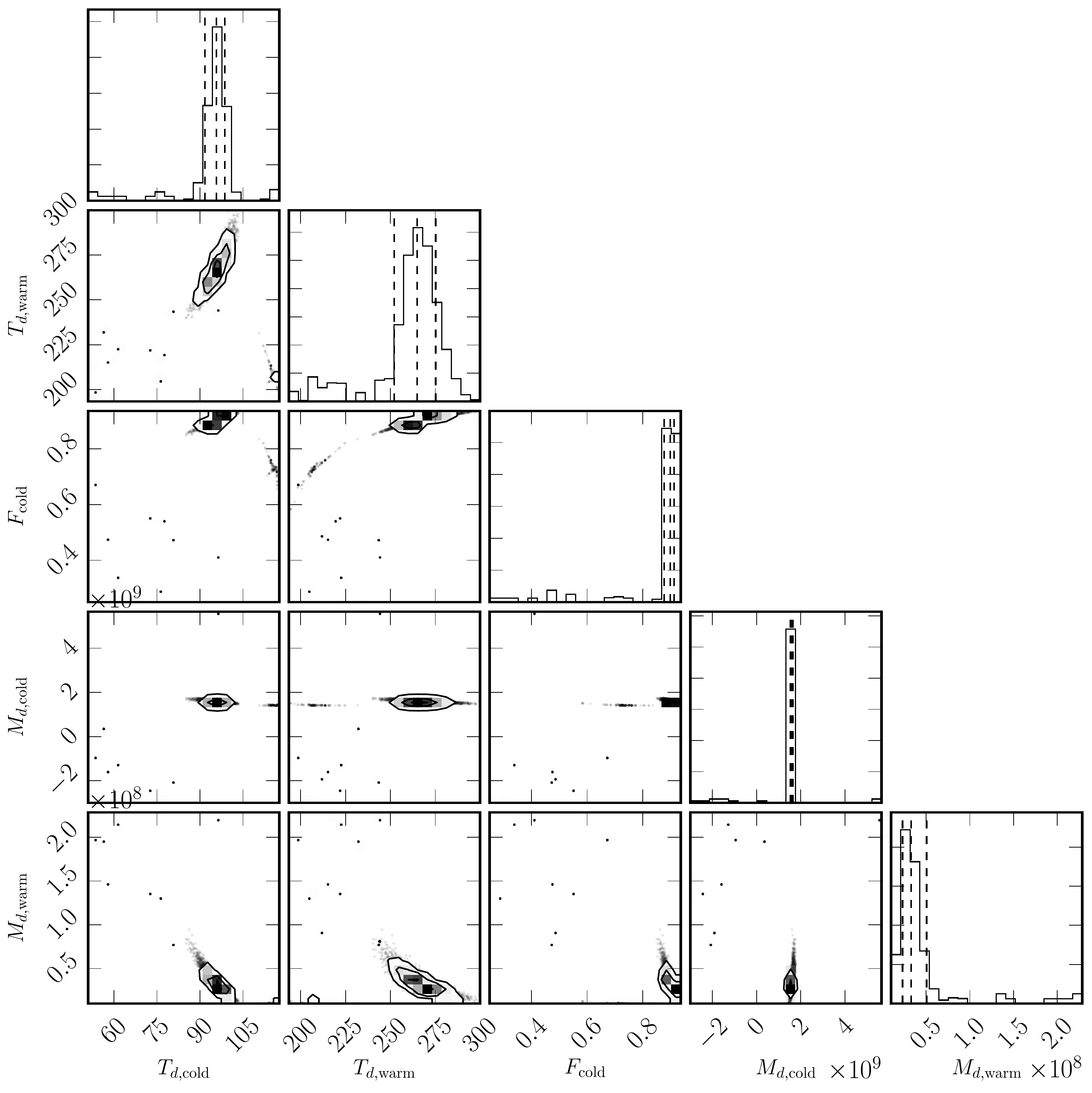}
\caption{2D correlation plots and 1D marginalized PDFs of the fully-analytical model (\Fig{corner_t1_t2_ff1_ff2_md1_md2}), with the vertical dashed line showing the 16th, 50th, and 84th percentiles.
\label{fig:corner1}}
\end{figure}

\begin{figure}[tbph]
\centering
\includegraphics[width=\textwidth]{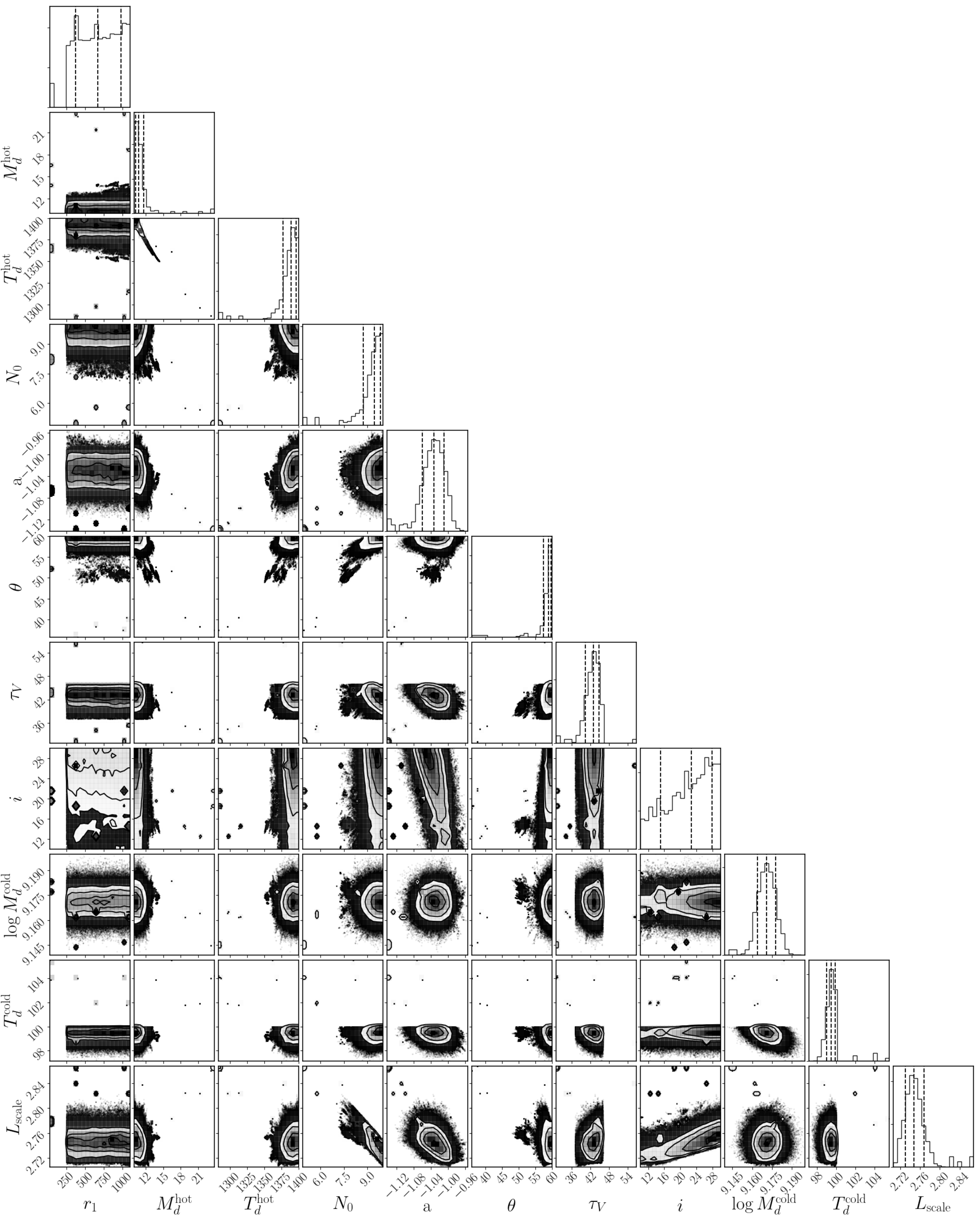}
\caption{2D correlation plots and 1D marginalized PDFs of the semi-analytical model (\Fig{torus}),
with the vertical dashed line showing the 16th, 50th, and 84th percentiles.
\label{fig:corner2}}
\end{figure}

\end{document}

%% file: table_photo.tex
\begin{deluxetable*}{lcccc}
\centering
\tabletypesize{\scriptsize}
\tablecolumns{5}
\tablecaption{Photometry data for \apm.}
\tablehead{
\colhead{Wavelength/Band} &
\colhead{Frequency} &
\colhead{Flux Density} &
\colhead{Instrument/Band}  &
\colhead{Reference}
\\
\colhead{($\micron$)} & \colhead{(GHz)} & \colhead{(mJy)} & \colhead{} & \colhead{} \vspace{0.05in}
}
\startdata
1.03 & 291060  & 5.00\pmm0.16          & {\it HST}/NICMOS     & 1 \\
1.25 & 239834  & 7.19\pmm0.20          & Keck I/NIRC          & 2 \\
1.55 & 193414  & 7.53\pmm0.19          & {\it HST}/NICMOS     & 1 \\
1.65 & 181692  & 8.97\pmm0.25          & Palomar/P200 InSb    & 2 \\
1.90 & 157786  & 7.81\pmm0.96          & {\it HST}/NICMOS     & 1 \\
2.15 & 139438  & 9.40\pmm0.26          & Palomar/P200 InSb    & 2  \\
3.4  & 88174   & 11.01\pmm0.23         & {\it WISE}/W1        & 3$^\dagger$  \\
3.5  & 85650   & 27.3\pmm1.0           & Palomar/P200 InSb    & 2 \\
3.8  & 79433   & 9.50\pmm1.22          & Subaru/IRCS          & 4 \\
3.6  & 83275   & 10.74\pmm0.0023          & {\em Spitzer}/IRAC   & 3$^\dagger$ \\ 
4.5  & 66620   & 7.88\pmm0.0021           & {\em Spitzer}/IRAC   & 3$^\dagger$ \\  
4.7  & 63994   & 7.20\pmm1.98         & Subaru/IRCS          & 4 \\
4.6  & 65172   & 10.86\pmm0.20         & {\it WISE}/W2        & 3$^\dagger$ \\
5.8  & 51688   &15.49\pmm0.007          & {\it Spitzer}/IRAC   & 3$^\dagger$ \\  
12   & 24983   & 62.92\pmm0.927        & {\it WISE}/W3        & 3$^\dagger$ \\
12   & 24983   & $<$101                & {\it IRAS}           & 5 \\
12.5 & 24000   & 74\pmm11              & Keck II/MIRLIN       & 2 \\
17.9 & 16750   & 103\pmm50             & Keck II/MIRLIN       & 2 \\
22   & 13626   & 187.6\pmm4.84         & {\it WISE}/W4        & 3$^\dagger$ \\
24   & 12491   & 186.8\pmm0.779        & {\it Spitzer}/MIPS   & 3$^\dagger$\\
25   & 12000   & 226.1\pmm16.2         & {\it IRAS}           & 5 \\
53   & 5656    & 567\pmm71             & {\it SOFIA}/HAWC$+$  & 3 \\
60   & 5000    & 511.1\pmm51.1         & {\it IRAS}           & 5 \\
70   & 4283    & 654.1\pmm8.745        & {\it Herschel}/PACS  & 3 \\
89   & 3368    & 682\pmm50             & {\it SOFIA}/HAWC$+$  & 3 \\
100  & 3000    & 951.1\pmm228          & {\it IRAS}           & 5 \\
154  & 1947    & 696\pmm43             & {\it SOFIA}/HAWC+    & 3 \\
160  & 1874    & 758.8\pmm13.2         & {\it Herschel}/PACS  & 3 \\
250  & 1200    & 627.2\pmm8.1          & {\it Herschel}/SPIRE & 3\\
350  & 857     & 431.4\pmm6.3          & {\it Herschel}/SPIRE & 3\\
350  & 850     & 386\pmm32             & CSO/SHARC II         & 6 \\
450  & 660     & 342\pmm26             & CSO/SHARC II         & 6 \\
450  & 660     & 285\pmm11             & JCMT/SCUBA           & 7 \\
450  & 660     & 211\pmm47             & JCMT/SCUBA           & 8 \\
500  & 600     & 249.6\pmm8.1          & {\it Herschel}/SPIRE & 3\\
850  & 350     & 84\pmm3               & JCMT/SCUBA           & 7 \\
850  & 350     & 75\pmm4               & JCMT/SCUBA           & 8 \\
1194 & 251     & 34\pmm0.55            & PdBI                 & 9  \\
1219 & 245.905 & 31.4\pmm2.0           & PdBI                 & 10  \\
1266 & 236.797 & 26.6\pmm1.3           & PdBI                 & 10  \\
1350 & 225     & 24\pmm2               & JCMT/SCUBA           & 8 \\
1400 & 214.0   & 17.0\pmm0.5           & PdBI                 & 11 \\
1400 & 211.1   & 16.9\pmm2.5           & PdBI                 & 12 \\
1490 & 201.166 & 16.5\pmm0.8           & PdBI                 & 10  \\
1958 & 153.132 & 5.4\pmm0.3            & PdBI                 & 10  \\
2710 & 110.7   & 2.17\pmm0.19          & CARMA                & 13  \\
2760 & 108.6   & 2.08\pmm0.22          & CARMA                & 13  \\
2840 & 105.4   & 2.04\pmm0.08          & CARMA                & 13  \\
3200 & 94.3    & 1.2\pmm0.3            & PdBI                 & 11 \\
3200 & 93.879  & 1.3\pmm0.2            & PdBI                 & 12 \\
3200 & 93.856  & 1.1\pmm0.02           & NOEMA                & 14 \\
3300 & 90.2    & 1.3\pmm0.2            & PdBI                 & 12 \\
3300 & 90.2    & 0.66\pmm0.18          & PdBI                 & 15 \\
7000 & 43.3    & $<$0.90               & VLA                  & 16 \\
13000 & 23.4649  & 0.30\pmm0.05        & VLA                  & 16 \\
13000 & 23.4649  & 0.41\pmm0.07        & VLA                  & 17 \\
13000   &   23.4649 &  0.38\pmm0.02    & VLA                  & 18    \\
20000   &   14.9399 &  0.30\pmm0.09    & VLA                  & 18    \\
36000   &   8.4601  &  0.45\pmm0.02    & VLA                  & 18    \\
36000   &   8.4601  &  0.45\pmm0.03    & VLA                  & 1    \\
60000   &   4.5276  &  0.55\pmm0.04    & VLA                  & 18   \\
200000  &   1.400   &  1.16\pmm0.03    & VLA                  & 18   \\
200000  &   1.400   &  0.92\pmm0.13    & VLA                  & 19\tna \\
900000  &   0.334   &  5.10\pmm0.80    & VLA                  & 20 \\
\enddata
\label{tab:photometry}
\tablecomments{
Uncertainties on the SPIRE flux densities include those due to confusion noise.
Uncertainties quoted here for the radio and mm interferometric measurements
do not include those from absolute flux calibration, but they are
accounted for in the SED modeling.
}
\tablenotetext{$\dagger$}{Available through the NASA/IRSA; first reported here. }
\tablenotetext{a}{VLA FIRST Catalog Database version 2014dec17.}
\tablerefs{
1: \citet{Ibata99a};
2: \citet{Egami00a};
3: This work;
4: \citet{Oya13a};
5: \citet{Irwin98a};
6: \citet{Beelen06a};
7: \citet{Barvainis02a};
8: \citet{Lewis98a};
9: \citet{Lis11a};
10: \citet{vdwerf11a};
11: \citet{Downes99a};
12: \citet{Weiss07a};
13: \citet{Riechers10a};
14: \citet{Feruglio17a};
15: \citet{Wagg05a};
16: \citet{Papadopoulos01a};
17: \citet{Lewis02a};
18: \citet{Riechers09a};
19: \citet{Becker94a};
20: \citet{Ivison06a}
}
\end{deluxetable*}

%% file: table_out.tex
\begin{deluxetable}{lcc}
\centering
\tabletypesize{\scriptsize}
\tablewidth{0.5\textwidth}
\tablecolumns{3}
\tablecaption{SED Model Parameters for \apm.}
\tablehead{
\colhead{Properties} &
\colhead{Units} &
\colhead{Values}
}
\startdata
\multicolumn{3}{c}{Analytical: Two-temperature MBB + Radio Power-laws\tna} \\  [.2em] \hline \noalign{\vskip .3em}
$T_{d, \rm warm}$         &  (K)          &      296\petm{17}{15}       \\  [.3em]
$T_{d, \rm cold}$         &  (K)          &      110\petm{3}{3}         \\  [.3em]
$F_{\rm Filling}^{\rm cold}$         &            &      0.9\petm{0.1}{0.1}         \\  [.3em]
$x_{\rm warm}$\tnb       &  (10$^{7}$\,\Msun\,pc$^{-2}$)  & 2.5\petm{1.2}{1.0}   \\  [.3em]
$x_{\rm cold}$\tnb       &  (10$^{9}$\,\Msun\,pc$^{-2}$)  & 1.4\petm{0.1}{0.1}   \\  [.3em]
$L_{\rm IR}$              &  (10$^{15}\mu$\pmOne\,\Lsun)   & 1.8\petm{0.7}{0.6}              \\  [.3em]
$L_{\rm FIR}$             &  (10$^{14}\mu$\pmOne\,\Lsun)   & 1.8\petm{0.1}{0.2}             \\
\cutinhead{Semi-analytical: Including Clumpy Torus\tnc} \\ [-1.5ex]
$r_{\rm hot}$           &          (pc)        & 655\petm{418}{409}  \\  [.3em]
$M_{d}^{\rm hot}$       &          ($\mu$\pmOne\,\Msun)  &  11\petm{5}{1} \\  [.3em]
$M_{d}^{\rm cold}$      &          (10$^9\mu$\pmOne\,\Msun)  &    1.4\petm{0.3}{0.2} \\  [.3em]
$T_d^{\rm hot}$         &          (K)       &  1385\petm{10}{49}     \\  [.3em]
$T_d^{\rm cold}$        &          (K)       &  100\petm{2}{1}       \\  [.3em]
$N_0$                   &                    & 8.3\petm{1.3}{2.2}  \\  [.3em]
$a$                     &                    & $-$1.0\petm{0.1}{0.1} \\  [.3em]
$\theta$                &        ($\degr$)   &   55\petm{4}{11}        \\  [.3em]
$\tau_V$                &                    &   42\petm{5}{5}      \\  [.3em]
$i$                     &        ($\degr$)   &   15\petm{8}{8}   \\  [.3em]
$\log{\left(L_{\rm scale}/\Lsun\right)}$   &         & 48.8\petm{0.1}{0.1} \\ [.3em]
$L_{\rm IR}$          &          (10$^{15}\mu$\pmOne\,\Lsun)          & 3.5\petm{0.1}{0.1}             \\  [.3em]
$L_{\rm NIR}$         &          (10$^{14}\mu$\pmOne\,\Lsun)    &       5.9\petm{0.4}{0.6}         \\  [.3em]
$L_{\rm MIR}$         &          (10$^{14}\mu$\pmOne\,\Lsun)          &       2.7\petm{0.3}{0.7}        \\  [.3em]
$L_{\rm FIR}$         &          (10$^{14}\mu$\pmOne\,\Lsun)          &       1.7\petm{0.5}{0.1}         \\  [.3em]
$L_{\rm FIR}^{\rm Torus}$         &          (10$^{13}\mu$\pmOne\,\Lsun)          &        4.7\petm{0.3}{0.5}     \\  [0.3em]      
$L_{\rm FIR}^{\rm Cold}$         &          (10$^{14}\mu$\pmOne\,\Lsun)          &        1.2\petm{0.5}{0.7}     \\  [-0.05in]      
\enddata
\label{tab:phy}
\tablecomments{
The different IR luminosities are obtained by integrating over rest-frame:
1$-$1000\,$\micron$ for $L_{\rm IR}$;
1$-$3\,$\micron$ for $L_{\rm NIR}$;
3$-$40\,$\micron$ for $L_{\rm MIR}$; and
42.5$-$122.5\,$\micron$ for $L_{\rm FIR}$.
}
\tablenotetext{a}{See \Sec{mbb}}
\tablenotetext{b}{$x\equiv M_d/r_0^2$, see Equation~\ref{eqn:tau}.}
\tablenotetext{c}{See \Sec{five}}
\end{deluxetable}